\UseRawInputEncoding
\documentclass[superscriptaddress, notitlepage, reprint,twocolumn]{revtex4-1}
\usepackage[english]{babel}
\usepackage{amsmath,amsthm,amssymb,lipsum}
\usepackage{amsfonts}
\usepackage[pdfborder={0 0 0}, colorlinks=true, urlcolor=blue, linkcolor=blue, citecolor=blue]{hyperref}
\usepackage{color}
\usepackage{graphicx}
\usepackage{float}
\usepackage{subfigure}
\usepackage{lipsum}
\usepackage{epstopdf}
\usepackage{dcolumn}
\usepackage{mathrsfs}
\begin{document}

\title{Dispersive Qubit Readout of Temperature}
\author{Dong  Xie}
\email{xiedong@mail.ustc.edu.cn}
\author{Chunling Xu}

\affiliation{College of Science, Guilin University of Aerospace Technology, Guilin, Guangxi 541004, People's Republic of China}

\begin{abstract}
 Squeezed light can exponentially increase the signal-to-noise ratio (SNR) of dispersive qubit readout, especially using a combination of injected external squeezing (IES) and intracavity squeezing (ICS). We further investigate whether IES and ICS can also exponentially improve the measurement precision of temperature. In the case of fully thermalized qubits isolated from thermal bath, the measurement precision of temperature can be improved exponentially when the temperature or measurement time or the input photon number approaches 0. In general, thermal fluctuations prevent the action of squeezed light. When multiple qubits maintain interacting with the thermal bath, the Heisenberg scaling can be achieved if the loss rate of the cavity is large and the coupling between the qubit and the optical cavity is weak enough. In the meantime, IES can also further promote the improvement of the measurement precision of the temperature exponentially.
\end{abstract}
\maketitle

\section{Introduction}
In the field of quantum communication\cite{lab1,lab2} and quantum error correction\cite{lab3}, a common and powerful approach is dispersive readout where a qubit is strongly detuned from a cavity mode. By measuring the output cavity field quadratures, i.e., homodyne detection, high-fidelity readout can be obtained.

In quantum metrology, squeezed light has been proved to improve the measurement sensitivity\cite{lab4}, such as, optomechanical motion sensing\cite{lab5,lab6}, and gravitational-wave
detection\cite{lab7,lab8}. In order to expand the application of squeezed light to dispersive qubit readout, Barzanjeh \emph{et al.} \cite{lab9} showed that an amplified interferometer can enhance the
quality of dispersive qubit measurement. However, as the dispersive interaction will lead to a frequency-dependent rotation of the squeezing axis, the corresponding amplified noise prevents  them from getting the Heisenberg
scaling. In order to obtain the true Heisenberg-limited scaling, Ref.\cite{lab10} has showed that injecting two-mode squeezed light into two coupled cavities can exponentially improve the SNR.
In the experiment\cite{lab11}, squeezed light has been used to reduce noise inside the signal to achieve higher SNR compared to that of coherent light readout in the same system. Recently, Qin \emph{et al.}\cite{lab12} and Kam \emph{et al.}\cite{lab13} proposed more suitable schemes of the standard dispersive readout, which used IES and ICS together to improve the SNR. Especially, Ref.\cite{lab12} has taken full advantage of
squeezing to obtain that the resulting SNR can scale as $e^{2r}$ for short-time measurements, rather than $e^r$ in Ref.\cite{lab10}, where $r$ refers to the squeezing parameter.

Precise measurement of temperature is an
important and significant subject in the field of quantum
metrology\cite{lab14,lab15,lab16,lab17,lab18}, quantum thermodynamics\cite{lab19} and quantum simulations\cite{lab20,lab21,lab22}. Since squeezed light can be used to facilitate dispersive qubit readout,  it is necessary to investigate whether IES and ICS can also exponentially improve the measurement precision of temperature.

In this article, we try to utilize the dispersive qubit to measure the temperature of a thermal bath. Firstly, we consider that the qubit is completely thermalized after passing through a thermal bath with the temperature $T$ and then weakly coupled to an optical cavity, and the temperature can be measured nondestructively by homodyne measurement. We find that the measurement precision of the temperature can be improved exponentially only when the measurement time tends to 0 or the temperature or the number of input photons goes to 0. Otherwise, neither IES or ICS can exponentially improve the measurement precision of the temperature. Then, we consider that $N$ qubits remain in contact with the thermal bath in the process of the dispersive measurement.
We show that the Heisenberg scaling can be obtained for small $N$ in the case of large loss rate of the
cavity and weak coupling between the qubits and the optical cavity. In the meantime, the measurement precision of the temperature can be improved exponentially by the squeezed light.
When the number of qubits is large enough, increasing the number of qubits and the squeezing parameter will reduce the  measurement precision of the temperature. No matter what the conditions are, increasing the number of input photons can always improve the measurement precision of the temperature.

The rest of article is ranged as follows. In section II, the physical setup of the dispersive qubit readout is introduced and the measurement uncertainty of the temperature is obtained with IES. In section III, IES and ICS are used simultaneously for dispersive qubit readout of the temperature. The standard quantum limit is recovered in section IV. In section V, the dispersive measurement is performed in the case of $N$ qubits continuously maintaining interaction with the thermal bath. We make a conclusion in section VI.

\section{dispersive measurement with IES}
\begin{figure}[h]
\includegraphics[scale=0.17]{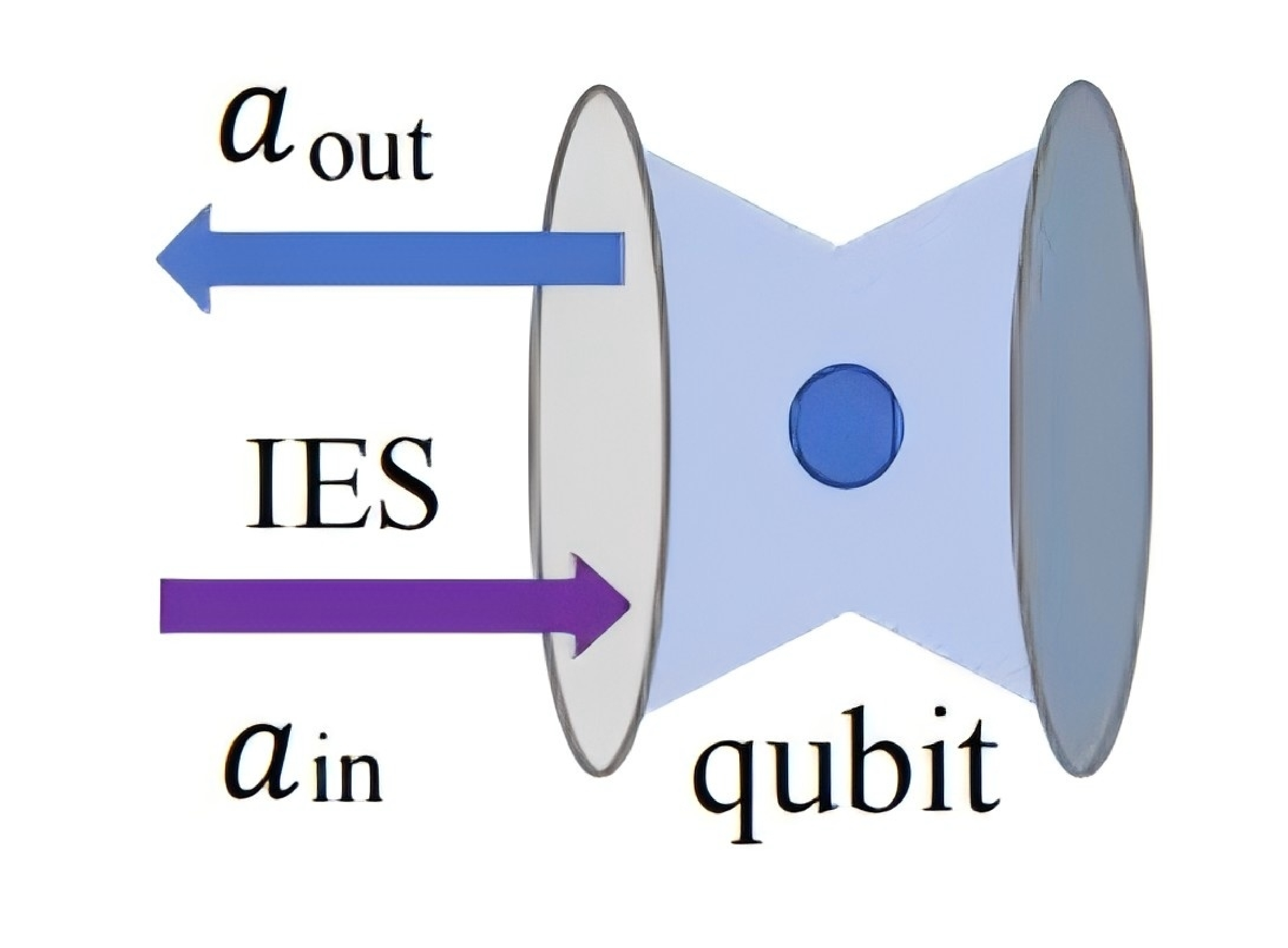}
 \caption{\label{fig.1} Diagram of dispersive qubit readout of temperature. The thermalized qubit with the temperature $T$ is placed in a cavity, which is driven by IES denoted as $a_{in}$. By homodyne detection of the output mode $a_{out}$, the temperature $T$  can be obtained.}
\end{figure}
We utilize a setup where a qubit dispersively couples to a cavity to measure the temperature of  a thermal bath, as shown in Fig.~\ref{fig.1}. The dispersive setup is dominated by the Hamiltonian ($\hbar=1$)\cite{lab23}
\begin{align}
{H}=\omega_c {a}^\dagger{a}+\frac{1}{2}\omega_q{\sigma}_z+\chi{\sigma}_z{a}^\dagger{a},
\end{align}
where $a$ $(a^\dagger)$ denote annihilation (creation) operators of the cavity mode with the frequency $\omega_c$, $\sigma_z$ is the Pauli matrix of the qubit with the transition frequency $\omega_q$, and $\chi$ is the coupling strength between the qubit and the cavity mode.
In a reference frame rotating at $\omega_c$, the quantum Langevin equation of the cavity mode ${a}$ can be described by
\begin{align}
\dot{{a}}=-i(-i\kappa/2+\chi{\sigma}_z){a} -\sqrt{\kappa}{a}_{in}(t),\label{eq:2}
\end{align}
where $\kappa$ denotes the cavity photon loss rate, and ${a}_{in}(t)$
represents the input field of the cavity.  Considering a squeezed vacuum reservoir, the correlations
for the input-noise operator ${A}_{in}(t)={a}_{in}(t)-\langle{a}_{in}(t)\rangle$ are given by
\begin{align}
\langle {A}_{in}(t){A}_{in}(t')\rangle=\frac{1}{2} e^{i\phi}\sinh(2r)\delta(t-t'),\label{eq:3}\\
\langle {A}^\dagger_{in}(t){A}^\dagger_{in}(t')\rangle=\frac{1}{2} e^{-i\phi}\sinh(2r)\delta(t-t'),\label{eq:4}\\
\langle {A}_{in}(t){A}^\dagger_{in}(t')\rangle=\cosh^2(r)\delta(t-t'),\label{eq:5}\\
\langle {A}^\dagger_{in}(t){A}_{in}(t')\rangle=\sinh^2(r)\delta(t-t'),\label{eq:6}
\end{align}
where $r$ is the squeezing parameter and $\phi$ denotes the reference phase.

Firstly, we consider that the thermalized qubit is isolated from the thermal bath after time $t=0$, so the system is always in thermal equilibrium
\begin{align}
 \rho_q=\frac{\exp[-\frac{\omega_q}{2T}\sigma_z]}{\exp[-\frac{\omega_q}{2T}]+\exp[\frac{\omega_q}{2T}]},\label{eq:7}
 \end{align}
 where we have already taken Boltzmann constant $\beta=1$, and the temperature $T$ of the thermal bath is what to be detected.

In the rotating reference frame, the solution of Eq.~(\ref{eq:2}) is analytically derived
\begin{align}
{a}(t)=\exp[\Lambda t]{a}(0) -\sqrt{\kappa}\int_0^tdt'\exp[\Lambda(t-t')]{a}_{in}(t'),\label{eq:8}
\end{align}
where the parameter $\Lambda=-i(-i\kappa/2+\chi{\sigma}_z)$.

The information of the temperature is encoded into output field of the cavity mode
\begin{align}
{a}_{out}(t)={a}_{in}(t)+\sqrt{\kappa}{a}(t).\label{eq:9}
\end{align}

For homodyne detection, the  corresponding output quadrature $Q$ is defined as
\begin{align}
{Q}(t)={a}_{out}(t) e^{-i\varphi}+{a}^\dagger_{out}(t) e^{i\varphi}\label{eq:10},
\end{align}
where $\varphi$ denotes the measurement angle.
 The essential parameter quantifying homodyne detection is the SNR, which is defined as\cite{lab24}
 \begin{align}
\textmd{SNR}=\frac{|\langle M\rangle_1-\langle M\rangle_0|}{\sqrt{\langle M_N^2\rangle_1+\langle M_N^2\rangle_0}}.\label{eq:11}
\end{align}
where the time-integrated output quadrature $M=\sqrt{\kappa}\int_0^\tau dt {Q}(t)$ with a measurement time $t$, the fluctuation noise operator $M_N=M-\langle M\rangle$,  and the lower corner marks 1 and 0 labels the excited and ground states of the qubit.
In the limits of $\kappa\tau\rightarrow0$ and $\kappa\tau\rightarrow\infty$, Ref. \cite{lab12} obtained an exponential decrease in the measurement noise
\begin{align}
\langle M_N^2\rangle_1+\langle M_N^2\rangle_0\simeq 2\kappa\tau e^{-2r}.\label{eq:12}
\end{align}
It means that the SNR is improved exponentially. Next, we investigate whether the measurement precision of the temperature can also be improved exponentially.

The error propagation function can be used to  quantify the measurement uncertainty of the temperature,
 \begin{align}
\delta T=\frac{\sqrt{\langle M_N^2\rangle}}{|\partial_T\langle M\rangle|}.\label{eq:13}
\end{align}

When the cavity mode is arriving at the steady state ($\tau\rightarrow \infty$) driven by a coherent input tone $\langle a_{in}\rangle=\alpha_{in} e^{i \theta}$,  we obtain the analytical form of the uncertainty for the conditions $\phi-2\varphi=\pi$ and $\arctan(2\chi/\kappa)=n\pi$,(see Appendix: A)
 \begin{align}
\delta T=\frac
{(\chi^2+\kappa^2/4)T^2(1+\cosh (\omega_q/T))\sqrt{e^{-2r}
+f(T)}}{2\alpha_{in}\kappa \sqrt{ \tau}\chi\omega_q},
\end{align}
where $f(T)=\frac{4\alpha^2_{in}\kappa^2\chi^2\tau(1-|\langle\sigma_z\rangle|^2)}{(\chi^2+\kappa^2/4)^2}$ represents the fluctuation of non-zero temperature, and $\langle \sigma_z\rangle=\frac{1-e^{\omega_q/T}}{1+e^{\omega_q/T}}$.
In this case, the precision of the temperature can not be improved exponentially unless the temperature $T$ or the input amplitude $\alpha_{in}$  goes to zero, which makes $f(T)=0$.

When $\tau\rightarrow0$, $\arctan(2\chi/\kappa)=n\pi$, and $\phi-2\varphi=\pi$, we can achieve
 \begin{align}
\delta T=\frac{e^{-r}(4\chi^2+\kappa^2)^2T^2(1+\cosh (\omega_q/T))}{4\alpha_{in}\kappa^4 \tau^{3/2}\chi\omega_q}.
\end{align}
From the above equation, we can see that the precision can be exponentially improved with the squeezing parameter $r$
in the limit of $\tau\rightarrow 0$. This is consistent with what we get with SNR.

\section{using IES and ICS together}
When IES and ICS are used simultaneously for dispersive qubit readout, it has been demonstrated that an exponential increase of the readout SNR can be obtained for
any measurement time \cite{lab12}. In this section, we explore whether the measurement precision of the temperature can also be exponentially improved at any time.

Including a two-photon driving of amplitude $\Omega$, frequency $\omega_d$, and phase $\theta'$, the total Hamiltonian in a reference frame rotating at $\omega_d$ is described as
\begin{align}
{H}_{tot}=\Delta_c {a}+\frac{1}{2}\Delta_q{\sigma}_z+\chi{\sigma}_z{a}^\dagger{a}+\Omega(e^{i \theta'}a^{\dagger2}+e^{-i \theta'}a^{2}),
\end{align}
where $\Delta_c=\omega_c-\frac{\omega_d}{2}$, and $\Delta_q=\omega_q-\frac{\omega_d}{2}$.
The corresponding quantum Langevin equation of the cavity mode ${a}$ can be described by
\begin{align}
\dot{{a}}=-i(\Delta_c-i\kappa/2+\chi{\sigma}_z){a}-i2\Omega e^{i\theta'}a^\dagger -\sqrt{\kappa}{a}_{in}(t).
\end{align}

In the case of $\Delta_c\neq0$, a Bogoliubov mode, $b=\cosh (r_c) a+e^{i\vartheta}\sinh (r_c) a^\dagger$, which is dominated by
\begin{align}
\dot{b}(t)=-i(\omega_{sq}-i\kappa/2+\chi_{sq}\sigma_z)b(t)-\sqrt{\kappa}b_{in}(t),\label{eq:18}
\end{align}
where $\omega_{sq}=\sqrt{\Delta_c^2-4\Omega^2}$ denotes the resonance frequency of the mode $b$, the the phase difference is $\vartheta=\theta-\varphi$, and $\chi_{sq}=\chi[\cosh(r_c)+\frac{\sinh^2(r_c)}{\cosh(r_c)+2\omega_{sq}\cosh(r_c)/(\Delta_q-\omega_{sq})}]$ denotes the dispersive coupling of the qubit and the mode $b$.  Here, the squeezing parameter $r_c$ satisfies that $\tanh (r_c)=2\Omega_{sq}/\Delta_c $.

Integrating the equation of motion in Eq.~(\ref{eq:18}), we obtain the analytical solution of the Bogoliubov mode $b$
\begin{align}
{b}(t)&=\exp[-i(\omega_{sq}-i\kappa/2+\chi_{sq}\sigma_z)t]b(0)-\nonumber\\
&\sqrt{\kappa}\int_0^tds\exp[-i(\omega_{sq}-i\kappa/2+\chi_{sq}\sigma_z)(t-s)]b_{in}(s).
\end{align}

According to the input-output relation, $b_{out}(t)=b_{in}(t)+\sqrt{\kappa}b(t)$ and taking the Bogoliubov transformation $a_{out}(t)=\cosh(r_c)b_{out}(t)-e^{i\theta'}\sinh(r_c)b^\dagger_{out}(t)$,
 we can obtain the value of $\langle M\rangle$ and $\langle M_N^2\rangle$.
When $r=r_c$, $\theta '-\phi=\pi$, and $\theta'=2\varphi=2\theta$, we obtain the measurement uncertainty of the temperature (see the detail in Appendix: B)
\begin{align}
\delta T=\frac{\sqrt{\nu^2(1-\langle\sigma_z\rangle^2)+\langle \delta M^2\rangle}}{|\nu\partial_T\langle\sigma_z\rangle|},
\end{align}
where $\langle \delta M^2\rangle=\kappa\tau e^{-2r}$. This result shows that the fluctuation from the non-zero temperature generally prevents the measurement precision from increasing exponentially.

When the system is close to the steady state, i.e., $\tau\gg1/\kappa$, we obtain the simplified form of $\nu$
  \begin{align}
\nu&=\frac{32\alpha_{in}\omega_{sq}\tau\chi_{sq}\kappa^{5/2}}{[\kappa^4+16(\omega_{sq}^2-\chi_{sq}^2)^2+8\kappa^2(\omega_{sq}^2+\chi_{sq}^2)]}.
 \end{align}

When $\tau\rightarrow 0$, the simplified form of $\nu$ is given by
 \begin{align}
\nu&=\frac{\alpha_{in}\kappa^{3/2}\omega_{sq}\chi_{sq}\tau^4}{6}.
 \end{align}

All the cases show that only when the temperature or the measurement time $\tau$ or the input photon number $\alpha_{in}$ is close to 0, the measurement precision can be improved exponentially by IES and ICS.

In contrast, when $\alpha_{in} \tau\rightarrow \infty$ or the squeezing
parameter $r\rightarrow \infty$, the optimal measurement precision of the temperature is obtained from the dispersive qubit readout,
\begin{align}
\delta T=\frac{\sqrt{(1-\langle\sigma_z\rangle^2)}}{|\partial_T\langle\sigma_z\rangle|}
=\frac{2T^2\sqrt{1+\cosh (\omega_q/T)}}{\omega_q }\label{eq:23}
\end{align}

According to the quantum Cram\'{e}r-Rao bound\cite{lab25,lab26,lab27}, the uncertainty of the temperature encoded into the matrix $\rho_q$ is given by
\begin{align}
\delta T\geq\frac{1}{\sqrt{\mathcal{F}(\rho_q)}},\label{eq:24}
\end{align}
where $\mathcal{F}(\rho_q)$ is the quantum Fisher information. Due to that the matrix $\rho_q$ is diagonal, the quantum Fisher information is calculated by
\begin{align}
\mathcal{F}(\rho_q)=\frac{(\partial_T P)^2}{P(1-P)},\label{eq:25}
\end{align}
where the probability that the qubit is in the ground state is given by
\begin{align}
P=\frac{\exp[-\frac{\omega_q}{2T}]}{\exp[-\frac{\omega_q}{2T}]+\exp[\frac{\omega_q}{2T}]}.
\label{eq:26}
\end{align}
Substituting Eq.~(\ref{eq:25}) and (\ref{eq:26}) into Eq.~(\ref{eq:24}), we get the optimal measurement precision of the temperature
\begin{align}
\delta T\geq\frac{2T^2\sqrt{1+\cosh (\omega_q/T)}}{\omega_q }\label{eq:27}.
\end{align}
Comparing Eq.~(\ref{eq:27}) obtained from the quantum Fisher information and Eq.~(\ref{eq:23}) obtained from the dispersive qubit readout, we can see that the temperature encoded into the qubit is completely obtained by the dispersive qubit readout in the limit case of  $\alpha_{in} \tau\rightarrow \infty$ or the squeezing
parameter $r\rightarrow \infty$.

\section{$N$ qubits }
Next, we consider using $N$ qubits to measure temperature. Including $N$ dispersive couplings between qubits and the cavity mode, the Hamiltonian is written as

\begin{align}
{H}_N=&\Delta_c {a}\nonumber\\
&+\sum_{j=1}^N(\frac{1}{2}\Delta_q{\sigma}_{jz}+\chi{\sigma}_{jz}{a}^\dagger{a})+\Omega(e^{i \theta'}a^{\dagger2}+e^{-i \theta'}a^{2}),
\end{align}
where ${\sigma}_{jz}$ denotes the Pauli operator of the $j$th qubit.
By the same process of calculation, we can obtain the measurement precision of the temperature. Here, we also consider that the $N$ qubits are isolated from the heat source after time $t = 0$. The initial states of qubits are always in thermal equilibrium
\begin{align}
 \rho_{Nq}=\frac{\prod_{j=1}^N\exp[-\omega_q/(2T)\sigma_{jz}]}{\exp[-\omega_q/(2T)]+\exp[\omega_q/(2T)]}.
 \end{align}

The optimal precision of the temperature is given by
\begin{align}
\delta T&=\frac{\sqrt{\langle(\sum_{j=1}^N\sigma_{jz})^2\rangle-\sum_{j=1}^N\langle\sigma_{jz}\rangle^2}}{|\partial_T\langle\sum_{j=1}^N\sigma_{jz}\rangle|}\\
&=\frac{\sqrt{(1-\langle\sigma_{jz}\rangle^2)}}{\sqrt{N}|\partial_T\langle \sigma_{jz}\rangle|}=\frac{2T^2\sqrt{1+\cosh (\omega_q/T)}}{\sqrt{N}\omega_q }.
\end{align}
Due to that $N$ qubits is independent, it only recovers the standard quantum limit.

\section{in contact with the thermal bath}
In order to obtain the Heisenberg limit, we consider that the $N$ qubits remain in contact with the thermal bath after reaching the thermal steady state. Measured in steady state,  using a combination of IES and ICS gives the same result as using IES alone. Therefore, we consider measuring the temperature without using ICS.
Assuming that the coupling strength of each qubit to the cavity is $\chi$, the Hamiltonian in the rotating reference frame is described by
\begin{align}
{H_{c}}=\sum_{i=1}^N\chi{\sigma}_{jz}{a}^\dagger{a}.
\end{align}

The quantum Langevin equation of the cavity mode ${a}$ and the qubits $\sigma_{jz}$ can be described by
\begin{align}
\dot{{a}}=&-i(-i\kappa/2+\sum_{i=1}^N\chi{\sigma}_{jz}){a} -\sqrt{\kappa}{a}_{in}(t),\\
\dot{{\sigma}_{jz}}=&-(4\Gamma n+2\Gamma){\sigma}_{jz}-2\Gamma+2\sqrt{2\Gamma}[\sigma^-\sigma_{jin}(t)\nonumber\\
&+\sigma^+\sigma^\dagger_{jin}(t)],
\end{align}
where $n=\frac{1}{\exp(\omega_q/T)-1}$ and the noise operators $\sigma_{jin}$ are characterized by the correlation functions\cite{lab28,lab29}
\begin{align}
\langle \sigma_{jin}(t)\sigma^\dagger_{jin}(t')\rangle&=({n}+1)\delta(t-t');\\
\langle \sigma^\dagger_{jin}(t)\sigma_{jin}(t')\rangle&={n}\delta(t-t').
\end{align}

The expectation values of the operators $a$ and $\sigma_{jz}$ in the steady state are given by
\begin{align}
\langle{a}\rangle&=\frac{\sqrt{\kappa}\alpha_{in}}{-iN\chi/(2n+1)+\kappa/2},\\
\langle{\sigma}_{jz}\rangle&=\frac{-1}{2n+1}.
\end{align}
Then, we get the measurement expectation value of the quadrature $Q=ae^{i\Phi}+a^\dagger e^{-i\Phi}$ in the steady state
\begin{align}
&\langle Q\rangle=\frac{\sqrt{\kappa}\alpha_{in} e^{i\Phi}}{-iN\chi/(2n+1)+\kappa/2}+\frac{\sqrt{\kappa}\alpha_{in} e^{-i\Phi}}{iN\chi/(2n+1)+\kappa/2}.
\end{align}
The signal obtained about the temperature $T$ is quantized as $S_T=|\partial_T\langle Q\rangle|$.
When $\Phi=\pi/2$, the maximal signal of the temperature $T$ is obtained
\begin{align}
&S^m_T=\frac{2\sqrt{\kappa}\alpha_{in} N\chi |\partial_Tn|(2n+1)}{N^2\chi^2+(2n+1)^2\kappa^2/4}.
\end{align}

To consider the effects of quantum fluctuations, we assume that $a=\langle a\rangle+\delta a$ and $\sigma_{jz}=\langle \sigma_{jz}\rangle+\delta \sigma_{jz}$. The corresponding quantum Langevin equation of the fluctuation operators $\delta a$ and $\delta \sigma_{jz}$ are described by
\begin{align}
\dot{\delta{a}}=-\kappa/2+iN\chi/(2n+1)\delta{a}-i\chi\sum_{i=1}^N \delta\sigma_{jz} -\sqrt{\kappa}{A}_{in}(t),\\
\dot{\delta{\sigma}_{jz}}=-(4\Gamma n+2\Gamma){\delta\sigma}_{jz}+2\sqrt{2\Gamma}[\sigma^-\sigma_{jin}(t)+\sigma^+\sigma^\dagger_{jin}(t)].
\end{align}

When $\Phi=\pi/2$ and $\kappa\gg\frac{2N\chi}{2n+1}e^r$, the uncertainty of the temperature is given by (see Appendix: C)
\begin{align}
&\delta T=\frac{\sqrt{\langle\Delta^2Q\rangle}}{S^m_T}\simeq\frac{(2n+1)\kappa^2e^{-r}}{8\sqrt{\kappa}\alpha_{in} N\chi |\partial_Tn|}.\label{eq:43}
\end{align}
This result shows that the Heisenberg scaling, i.e., $\delta T\propto 1/N$, has been achieved in the case of the large loss rate $\kappa$ and the weak coupling $\chi$. The intuitive explanation is that in the case of fast dissipation, the cavity system reaches the stable state faster than the qubits, and then collects the temperature information transmitted by the $N$ qubits. Due to the weak coupling $\chi$, the memory effect of the qubits can be negligible. It leads to the coherent accumulation of the temperature information in the state of the cavity mode.
In this case, ICS can also further promote the improvement of the precision exponentially.

When $\kappa\ll\frac{2N\chi}{2n+1}$ and $4n\Gamma+2\Gamma\ll\frac{2N\chi}{2n+1}$, the uncertainty is given by
\begin{align}
&\delta T=\frac{N\chi\sqrt{8(2n+1)(2n^2+4n+1)\Gamma+2\kappa \cosh (2r)}}{8{\kappa}\alpha_{in}  |\partial_Tn|(2n+1)}.\label{eq:44}
\end{align}

The above eqiuation shows that the measurement precision of the temperature will reduce with the number of qubits and the squeezing parameter.  This is because the memory of qubits becomes non-negligible when the total coupling strength is sufficiently strong by increasing the number of qubits and enhancing the squeezing parameter. As a  result, the temperature information can not be accumulated on the cavity mode.
\begin{figure}[h]
\includegraphics[scale=0.7]{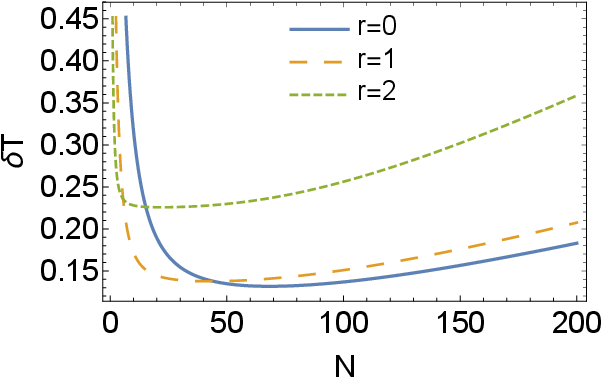}
 \caption{\label{fig.2} The graph of the uncertainty of temperature $\delta T$ varying with the number of qubits $N$. Three different lines represent different scenarios of squeezing parameters. Here, the dimensionless parameters are chosen as: $\kappa=100$, $T=1$, $\omega_q=1$, $\chi=1$, $\Gamma=10$, and $\alpha=100$.}
\end{figure}

The general correlation between the uncertainty of temperature $\delta T$ and the number of qubits $N$ is shown in Fig.~\ref{fig.2}. We can see that the uncertainty of temperature $\delta T$ decreases with the number of qubits $N$ when $N$ is less than a certain value. Conversely, when $N$ is greater than a certain value, the measurement certainty $\delta T$ gradually increases linearly with $N$. When $N$ is less than a certain value, the uncertainty of temperature $\delta T$ decreases with the squeezing parameter $r$ exponentially; When $N$ is greater than a certain value, the measurement certainty $\delta T$ increases with $r$.
 This is consistent with the results obtained in extreme cases as shown in Eq.~(\ref{eq:43}) and (\ref{eq:44}).

\section{conclusion}
In this article, we have investigated the measurement precision obtained by the dispersive qubit readout. We mainly explored two different cases: One is that the qubit is thermalized by the thermal bath and no longer has any interaction with the thermal bath; The second is that qubits are constantly interacting with the thermal bath. In the first case,  whether IES and ICS are used alone or together, the measurement precision of the temperature is generally not increased exponentially unless the temperature or the
input photon number or the measurement time approaches 0. Therefore, in general, this result is inconsistent with the conclusions obtained by the SNR, which is used to quantified the homodyne detection. This is mainly due to thermal fluctuations that prevent the action of squeezed light. In the second case, we obtain the Heisenberg scaling when the dissipation rate
is large enough and the coupling between the cavity and the qubits is weak enough. In this case, the measurement precision of the temperature can be improved exponentially by IES.
Our study will help to obtain high measurement precision of temperature based on the dispersive qubit readout, which is feasible in the field of current quantum technology\cite{lab30}.
\section*{Acknowledgements}
This research was supported by the National Natural Science Foundation of China (Grant No. 12365001 and No. 62001134), the Bagui youth top talent training program, and Guangxi Natural Science Foundation (Grant No. 2020GXNSFAA159047).
\newpage
\section*{Appendix: A}
In this section, we derive the general form of the measurement uncertainty of the temperature when only IES is used. Based on this, the results in the limit case can be obtained relatively easily.
Considering a coherent measurement tone $\langle a_{in}\rangle=\alpha_{in} e^{i \theta}$, we can obtain the expectation value of the measurement operator $M$ by using Eq.~(\ref{eq:3})-(\ref{eq:10})
 \begin{align}
&\langle M\rangle=[\sqrt{\kappa}2\alpha_{in}\tau e^{i\vartheta}-\frac{\kappa \alpha_{in}(1-e^{\Lambda \tau}+\Lambda \tau)}{\Lambda^2}e^{i\vartheta}]+H.c.\nonumber\\
&=\sqrt{\kappa}4\alpha_{in}\tau \cos (\vartheta)+
\frac{\kappa^{3/2}\alpha_{in}[2A(\chi^2-\kappa^2/4)-2B\chi\kappa]\cos (\vartheta)}{[\chi^2+(\kappa/2)^2]^2}\nonumber\\
&-\frac{\kappa^{3/2}\alpha_{in}[2A\kappa\chi+2B(\chi^2-\kappa^2/4)]\sin (\vartheta) \langle\sigma_z\rangle}{[\chi^2+(\kappa/2)^2]^2}\tag{A1}
\end{align}
where the parameters $\vartheta=\theta-\varphi$, $A=1-\kappa\tau/2-e^{-\kappa\tau/2}\cos(\chi\tau)$ and $B=e^{-\kappa\tau/2}\sin(\chi\tau)-\chi\tau$.

By utilizing the correlations for the input-noise operators and these initial conditions, we obtain the measurement noise
 \begin{align}
\langle M_N^2\rangle=\mu^2(1-\langle\sigma_z\rangle^2)+\langle \delta M^2\rangle\tag{A2}
\end{align}
where $\mu=\frac{\kappa^{3/2}\alpha_{in}[2A\kappa\chi+2B(\chi^2-\kappa^2/4)]\sin (\phi) }{[\chi^2+(\kappa/2)^2]^2}$, and $\langle \sigma_z\rangle=\frac{1-e^{\omega_q/T}}{1+e^{\omega_q/T}}$.
And $\langle \delta M^2\rangle$ denotes the noise term outside of thermal fluctuations, which is specifically stated as
\begin{widetext}
 \begin{align}
\langle \delta M^2\rangle&=\kappa\tau\cosh(2r)+\frac{\sinh(2r)}{2}\{3\cos(\phi-2\varphi)-
(3-2\kappa\tau)[\cos(2\varphi-\phi)\cos(4\psi)+\sin(4\psi)\sin(2\varphi-\phi)\langle\sigma_z\rangle]\nonumber\\
&-16e^{-\kappa\tau/2}\cos (\psi)\sin(2\psi)[\cos(3\psi)\sin(2\varphi-\phi+\chi\tau)\langle\sigma_z\rangle
+\sin(3\psi)\cos(2\varphi-\phi+\chi\tau)]\nonumber\\
&+4e^{-\kappa\tau}\cos(\psi)\sin(2\psi)[\cos(3\psi)\sin(2\varphi-\phi+2\chi\tau)\langle \sigma_z\rangle+\sin(3\psi)\cos(2\varphi-\phi+2\chi\tau)]\nonumber\\
&+6\sin(2\psi)\cos(2\varphi-\phi)\sin(4\psi)+6\sin(2\psi)\sin(2\varphi-\phi)\cos(4\psi)\langle \sigma_z \rangle\},\tag{A3}
\end{align}
\end{widetext}
where $\tan(\psi)=2\chi/\kappa$.
By using the error propagation function, the uncertainty of the temperature is given by

\begin{align}
\delta T=\frac{\sqrt{\langle M_N^2\rangle}}{|\partial_T\langle M\rangle|}
=\frac{\sqrt{\mu^2(1-\langle\sigma_z\rangle^2)+\langle \delta M^2\rangle}}{|\mu\partial_T\langle\sigma_z\rangle|}.\tag{A4}
\end{align}

When the cavity mode is arriving at the steady state ($\tau\rightarrow \infty$), and $\phi-2\varphi=\pi$, we obtain the analytical form of the uncertainty of the temperature
 \begin{align}
\delta T=\frac
{\sqrt{\cosh(2r)-\sinh(2r)\cos(4\psi)
+\frac{4\alpha^2_{in}\kappa^2\chi^2\tau(1-|\langle\sigma_z\rangle|^2)}{(\chi^2+\kappa^2/4)^2}}}{2\alpha_{in}\kappa \sqrt{ \tau}\chi|\partial_T\langle\sigma_z\rangle|/(\chi^2+\kappa^2/4)}\tag{A5}
\end{align}
When $\psi=n\pi$, the uncertainty is further simplified as
 \begin{align}
\delta T& =\frac
{\sqrt{e^{-2r}
+\frac{4\alpha^2_{in}\kappa^2\chi^2\tau(1-|\langle\sigma_z\rangle|^2)}{(\chi^2+\kappa^2/4)^2}}}{2\alpha_{in}\kappa \sqrt{ \tau}\chi|\partial_T\langle\sigma_z\rangle|/(\chi^2+\kappa^2/4)}\tag{A6}\\
&=\frac
{(\chi^2+\kappa^2/4)T^2(1+\cosh (\omega_q/T))\sqrt{e^{-2r}
+f(T)}}{2\alpha_{in}\kappa \sqrt{ \tau}\chi\omega_q},\tag{A7}
\end{align}
where $f(T)=\frac{4\alpha^2_{in}\kappa^2\chi^2\tau(1-|\langle\sigma_z\rangle|^2)}{(\chi^2+\kappa^2/4)^2}$.

In the limt of $\tau\rightarrow0$, we get that
 \begin{align}
\delta T=\frac{\sqrt{\delta+
16\alpha_{in}^2\kappa^8 \tau^3\chi^2(1-\langle\sigma_z\rangle^2)/(4\chi^2+\kappa^2)^4}}{4\alpha_{in}\kappa^4 \tau^{3/2}\chi|\partial_T\langle\sigma_z\rangle|/(4\chi^2+\kappa^2)^2},\tag{A8}
\end{align}
in which,
\begin{align}
\delta=&\cosh(2r)-\sinh(2r)[\cos(4\psi)+\cos(\psi)\sin(2\psi)\sin(3\psi)]\nonumber\\
&+4\chi\cos (\psi)\cos (3\psi)\sin (\psi)\langle \sigma_z\rangle/\kappa.\tag{A9}
\end{align}
When $\psi=n\pi$ and $\tau\rightarrow0$, we can achieve the uncertainty of the temperature
 \begin{align}
\delta T&=\frac{e^{-r}(4\chi^2+\kappa^2)^2}{4\alpha\kappa^4 \tau^{3/2}\chi|\partial_T\langle\sigma_z\rangle|}\tag{A10}\\
&=\frac{e^{-r}(4\chi^2+\kappa^2)^2T^2(1+\cosh (\omega_q/T))}{4\alpha_{in}\kappa^4 \tau^{3/2}\chi\omega_q}.\tag{A11}
\end{align}
\section*{Appendix: B}
In this section, we obtain the general formula of the measurement uncertainty of the temperature when IES and ICS are used simultaneously for dispersive qubit readout.

According to the input-output relation, $b_{out}(t)=b_{in}(t)+\sqrt{\kappa}b(t)$ and taking the Bogoliubov transformation $a_{out}(t)=\cosh(r_c)b_{out}(t)-e^{i\theta'}\sinh(r_c)b^\dagger_{out}(t)$,
 we can obtain the value of $\langle M\rangle$. Here, $\tanh (r_c)=2\Omega_{sq}/\Delta_c $.
When $r=r_c$, $\theta '-\phi=\pi$, and $\theta'=2\varphi=2\theta$,
we derive that

\begin{widetext}
  \begin{align}
&\langle M\rangle=\nonumber\\
&2\alpha_{in}\sqrt{\kappa}
\{[16(\omega_{sq}^2-\chi_{sq}^2)^2-\kappa^4][\kappa^4\tau-4\kappa^3+16\tau(\omega_{sq}^2-\chi_{sq}^2)^2-16\kappa(\omega_{sq}^2+\chi_{sq}^2)+8\kappa^2\tau(\omega_{sq}^2+\chi_{sq}^2)]-4e^{-\kappa\tau/2}\kappa[(4\omega_{sq}^2\nonumber\\
&+\kappa^2)^2-16\chi_{sq}^4]\cos(\chi_{sq}\tau)[(\kappa^2-4\omega_{sq}^2+4\chi_{sq}^2)\cos(\omega_{sq}\tau)-4\kappa\omega_{sq} \sin(\omega_{sq}\tau)]+16e^{-\kappa\tau/2}\kappa\chi_{sq}[32e^{\kappa\tau/2}\kappa^2\omega_{sq}^2\chi_{sq}-32\kappa^2\nonumber\\
&\omega_{sq}\chi_{sq}^2\cos(\chi_{sq}\tau)\cos(\omega_{sq}\tau)]-8\kappa\omega_{sq}\chi_{sq}\cos(\chi_{sq}\tau)\sin(\omega_{sq}\tau)(\kappa^2-4\omega_{sq}^2+4\chi_{sq}^2)+\sin(\chi_{sq}\tau)\cos(\omega_{sq}\tau)\kappa(\kappa^4+8\kappa^2\chi_{sq}^2-\nonumber\\
&8\kappa^2\omega_{sq}^2+16\chi_{sq}^4+32\omega_{sq}^2\chi_{sq}^2-48\omega_{sq}^4)-2\omega_{sq}\sin(\chi_{sq}\tau)\sin(\omega_{sq}\tau)(3\kappa^4-16\omega_{sq}^4-16\chi_{sq}^4+32\omega_{sq}^2\chi_{sq}^2+8\kappa^2\omega_{sq}^2+8\kappa^2\chi_{sq}^2)\nonumber\\
&+4\kappa\langle\sigma_z\rangle e^{-\kappa\tau/2}[4e^{\kappa\tau/2}\omega_{sq}\chi_{sq}(\kappa^2+4\omega_{sq}^2)(\kappa^3\tau-6\kappa^2+8\omega_{sq}^2+4\kappa\tau\omega_{sq}^2)+32e^{\kappa\tau/2}\omega_{sq}\chi_{sq}^3(\kappa^3\tau-2\kappa^2-8\omega_{sq}^2-4\kappa\tau\omega_{sq}^2)\nonumber\\
&+64e^{\kappa\tau/2}\omega_{sq}\chi_{sq}^5(2+\kappa\tau)+8\omega_{sq}\chi_{sq}\cos(\omega_{sq}\tau)\cos(\chi_{sq}\tau)(6\chi_{sq}\kappa^4-16\omega_{sq}^4-16\chi_{sq}^4+32\omega_{sq}^2\chi_{sq}^2+8\kappa^2\omega_{sq}^2+8\kappa^2\chi_{sq}^2)\nonumber\\
&+4\omega_{sq}\kappa\cos(\omega_{sq}\tau)\sin(\chi_{sq}\tau)(\kappa^4+16\omega_{sq}^4+8\kappa^2\omega_{sq}^2-8\kappa^2\chi_{sq}^2+32\omega_{sq}^2\chi_{sq}^2-48\chi_{sq}^4)+4\kappa\chi_{sq}\sin(\omega_{sq}\tau)\cos(\chi_{sq}\tau)(\kappa^4\nonumber\\
&+8\kappa^2\chi_{sq}^2-8\kappa^2\omega_{sq}^2+16\chi_{sq}^4+32\omega_{sq}^2\chi_{sq}^2-48\omega_{sq}^4)+\sin(\omega_{sq}\tau)\sin(\chi_{sq}\tau)(\kappa^6+4\kappa^4\omega_{sq}^2+4\kappa^4\chi_{sq}^2-64\omega_{sq}^6-64\chi_{sq}^6\nonumber\\
&+64\omega_{sq}^2\chi_{sq}^4+64\omega_{sq}^4\chi_{sq}^2-16\kappa^2\omega_{sq}^4-16\kappa^2\chi_{sq}^4+160\kappa^2\chi_{sq}^2\omega_{sq}^2)]\}/[\kappa^4+16(\omega_{sq}^2-\chi_{sq}^2)^2+8\kappa^2(\omega_{sq}^2+\chi_{sq}^2)]^2.
 \tag{B1} \end{align}
\end{widetext}
The measurement noise is given by
\begin{align}
\langle M_N^2\rangle=\nu^2(1-\langle\sigma_z\rangle^2)+\langle \delta M^2\rangle,\tag{B2}
\end{align}
where $\nu$ is described by
\begin{widetext}
\begin{align}
\nu&=8\alpha_{in}\kappa^{3/2} e^{-\kappa\tau/2}[4e^{\kappa\tau/2}\omega_{sq}\chi_{sq}(\kappa^2+4\omega_{sq}^2)(\kappa^3\tau-6\kappa^2+8\omega_{sq}^2+4\kappa\tau\omega_{sq}^2)+32e^{\kappa\tau/2}\omega_{sq}\chi_{sq}^3(\kappa^3\tau-2\kappa^2-8\omega_{sq}^2-4\kappa\tau\omega_{sq}^2)\nonumber\\
&+64e^{\kappa\tau/2}\omega_{sq}\chi_{sq}^5(2+\kappa\tau)+8\omega_{sq}\chi_{sq}\cos(\omega_{sq}\tau)\cos(\chi_{sq}\tau)(6\chi_{sq}\kappa^4-16\omega_{sq}^4-16\chi_{sq}^4+32\omega_{sq}^2\chi_{sq}^2+8\kappa^2\omega_{sq}^2+8\kappa^2\chi_{sq}^2)\nonumber\\
&+4\omega_{sq}\kappa\cos(\omega_{sq}\tau)\sin(\chi_{sq}\tau)(\kappa^4+16\omega_{sq}^4+8\kappa^2\omega_{sq}^2-8\kappa^2\chi_{sq}^2+32\omega_{sq}^2\chi_{sq}^2-48\chi_{sq}^4)+4\kappa\chi_{sq}\sin(\omega_{sq}\tau)\cos(\chi_{sq}\tau)(\kappa^4\nonumber\\
&+8\kappa^2\chi_{sq}^2-8\kappa^2\omega_{sq}^2+16\chi_{sq}^4+32\omega_{sq}^2\chi_{sq}^2-48\omega_{sq}^4)+\sin(\omega_{sq}\tau)\sin(\chi_{sq}\tau)(\kappa^6+4\kappa^4\omega_{sq}^2+4\kappa^4\chi_{sq}^2-64\omega_{sq}^6-64\chi_{sq}^6\nonumber\\
&+64\omega_{sq}^2\chi_{sq}^4+64\omega_{sq}^4\chi_{sq}^2-16\kappa^2\omega_{sq}^4-16\kappa^2\chi_{sq}^4+160\kappa^2\chi_{sq}^2\omega_{sq}^2)]/[\kappa^4+16(\omega_{sq}^2-\chi_{sq}^2)^2+8\kappa^2(\omega_{sq}^2+\chi_{sq}^2)]^2.\tag{B3}
 \end{align}

\end{widetext}
and  $\langle \delta M^2\rangle=\kappa\tau e^{-2r}$.
\section*{Appendix: C}
To make it easier to read, let us rewrite the quantum Langevin equation of the linear operators $\delta a$ and $\delta \sigma_{jz}$
\begin{align}
\dot{\delta{a}}=-\kappa/2+iN\chi/(2n+1)\delta{a}-i\chi\sum_{i=1}^N \delta\sigma_{jz} -\sqrt{\kappa}{A}_{in}(t),\tag{C1}\\
\dot{\delta{\sigma}_{jz}}=-(4\Gamma n+2\Gamma){\delta\sigma}_{jz}+2\sqrt{2\Gamma}[\sigma^-\sigma_{jin}(t)+\sigma^+\sigma^\dagger_{jin}(t)],\tag{C2}
\end{align}
where $\sigma^-=|0\rangle\langle1|$ and $\sigma^+=|1\rangle\langle0|$.
By integrating the above equations of motion, we obtain the analytical solutions
\begin{align}
&\delta{a}(t)=\exp[-i(-i\kappa/2-\frac{N\chi}{(2n+1)})t]\delta{a}(0) +\int_{0}^t\nonumber\\
&\exp[(-\kappa/2+\frac{iN\chi}{2n+1})(t-s)](-iN\chi\delta\sigma_{jz}(s) -\sqrt{\kappa}{A}_{in}(s)),\tag{C3}\\
&{\delta{\sigma}_{jz}}(t)=\exp[-(4\Gamma n+2\Gamma)t]{\delta\sigma}_{jz}(0)+\nonumber\\
&\int_0^tds2\sqrt{2\Gamma}\exp[-(4\Gamma n+2\Gamma)(t-s)]\sigma_{zin}(s),\tag{C4}
\end{align}
where the noise operators $\sigma_{zin}(t)=\sigma^-\sigma_{jin}(t)+\sigma^+\sigma^\dagger_{jin}(t)$. It satisfies the correlation
\begin{align}
\langle\sigma_{zin}(t)\sigma_{zin}(t')\rangle\simeq[1+n+n/(1+2n)]\delta(t-t'),\tag{C5}
\end{align}
where we use the mean field approximation $\langle\sigma^-\sigma^+\sigma_{jin}(t)\sigma^\dagger_{jin}(t')\rangle\simeq\langle \sigma^-\sigma^+\rangle\langle \sigma_{jin}(t)\sigma^\dagger_{jin}(t')\rangle$.

For the steady state, $\tau\gg1/\kappa$ and $\tau\gg1/(4n\Gamma+2\Gamma)$, the fluctuation operator of the cavity operator is rewritten as
\begin{align}
&\delta{a}_s=\nonumber\\
&\lim_{t\rightarrow\infty} -\sqrt{\kappa}\int_0^tds\exp[(-\kappa/2+\frac{iN\chi}{2n+1})(t-s)]A_{in}(s)ds\nonumber\\
&-2i\sqrt{2\Gamma}N\chi\int_0^tds\int_0^sds'\exp[(-\kappa/2+\frac{iN\chi}{2n+1})(t-s)\nonumber\\
&-(4\Gamma n+2\Gamma)(s-s')]\sigma_{zin}(s').\tag{C6}
\end{align}

The expectation values of the correlated fluctuation operators are derived by using the correlations of the noise operators $A_{in}$ and $\sigma_{zin}$,
\begin{align}
\langle(\delta{a}_s)^2\rangle&=\frac{\kappa e^{i \phi}}{2(\kappa-\frac{2iN\chi}{2n+1})}\sinh (2r)+\nonumber\\
&\frac{-2N^2\chi^2(2n^2+4n+1)}{(2n+1)^2(\kappa/2-\frac{iN\chi}{2n+1})(\kappa/2-\frac{iN\chi}{2n+1}+4\Gamma n+2\Gamma)},\tag{C7}\\
\langle\delta{a}_s^\dagger\delta{a}_s\rangle&={\sinh^2 (r)}\nonumber\\\
&+\frac{4N^2\chi^2(\kappa/2+4\Gamma n+2\Gamma)(2n^2+4n+1)}{\kappa(2 n+1)^2[\frac{N^2\chi^2}{(2n+1)^2}+(\kappa/2+4\Gamma n+2\Gamma)^2]}.\tag{C8}
\end{align}

Then, we get the measurement expectation value of the quadrature $Q=ae^{i\Phi}+a^\dagger e^{-i\Phi}$ in the steady state

\begin{align}
&\langle Q\rangle=\frac{\sqrt{\kappa}\alpha_{in} e^{i\Phi}}{-iN\chi/(2n+1)+\kappa/2}+\frac{\sqrt{\kappa}\alpha_{in} e^{-i\Phi}}{iN\chi/(2n+1)+\kappa/2}.\tag{C9}
\end{align}

The signal obtained about $T$ is quantized as $S_T=|\partial_T\langle Q\rangle|$.
When $\Phi=\pi/2$, the maximal signal of $T$ is given by
\begin{align}
&S^m_T=\frac{\sqrt{\kappa}\alpha_{in} N2\chi |\partial_Tn|(2n+1)}{N^2\chi^2+(2n+1)^2\kappa^2/4},\tag{C10}
\end{align}
where the partial derivative $|\partial_Tn|=(n^2+n)\omega/T^2$.
When $\Phi=\pi/2$, the variance $\langle\Delta^2Q\rangle=\langle Q^2\rangle-|\langle Q\rangle|^2$ is further described by
\begin{align}
&\langle\Delta^2Q\rangle=2\langle\delta{a}_s^\dagger\delta{a}_s\rangle+1-\langle(\delta{a}_s)^2\rangle-\langle(\delta{a}^\dagger_s)^2\rangle.\tag{C11}
\end{align}
Based on the error propagation formula in Eq.~(\ref{eq:13}) in the main text, the uncertainty of the temperature is derived by
\begin{align}
&\delta T=\frac{\sqrt{\langle\Delta^2Q\rangle}}{S^m_T}=\frac{\sqrt{2\langle\delta{a}_s^\dagger\delta{a}_s\rangle+1-\langle(\delta{a}_s)^2\rangle-\langle(\delta{a}^\dagger_s)^2\rangle}}{S^m_T}.\tag{C12}
\end{align}

When the loss rate is larger than the number of the qubits and the squeezing parameter, i.e., $\kappa\gg\frac{2N\chi}{2n+1}e^r$, the uncertainty of the temperature is given by
\begin{align}
&\delta T\simeq\frac{(2n+1)\kappa^2e^{-r}}{8\sqrt{\kappa}\alpha_{in} N\chi |\partial_Tn|}.\tag{C13}
\end{align}
When $\kappa\ll\frac{2N\chi}{2n+1}$ and $4n\Gamma+2\Gamma\ll\frac{2N\chi}{2n+1}$, the uncertainty is given by
\begin{align}
&\delta T=\frac{N\chi\sqrt{8(2n+1)(2n^2+4n+1)\Gamma+2\kappa \cosh (2r)}}{8{\kappa}\alpha_{in}  |\partial_Tn|(2n+1)}.\tag{C14}
\end{align}

\end{document}